\documentclass[vecphys]{svmult}

\usepackage{makeidx}         
\usepackage{graphicx}        
\usepackage{multicol}        
\usepackage[bottom]{footmisc}

\makeindex             

\begin{document}

\title*{The impact of  encounters on the members of Local Group 
Analogs. A view from {\it GALEX}}
\titlerunning{Local Group Analogs} 
\author{Buson, L. M.\inst{1},
Bettoni, D.\inst{1}, Bianchi, L.\inst{2}, Buzzoni, A.\inst{3}, Marino, A.\inst{1}
\and Rampazzo, R.\inst{1}}
\authorrunning{Buson et al.} 
\institute{INAF - Osservatorio Astronomico di Padova, Vicolo dell'Osservatorio 5, 
I-35122 Padova, Italy \texttt{lucio.buson@oapd.inaf.it, daniela.bettoni@oapd.inaf.it, 
antonietta.marino@oapd.inaf.it, roberto.rampazzo@oapd.inaf.it} \and
Department of Physics and Astronomy, Johns Hopkins University, 3400 North Charles 
Street, Baltimore, MD 21218 \texttt{bianchi@pha.jhu.edu}
\and INAF - Osservatorio Astronomico di Bologna, Via Ranzani 1, I-40127 Bologna, Italy
\texttt{alberto.buzzoni@oabo.inaf.it}}
%
%
\maketitle

\begin{abstract}
The bright galaxy population of the Local Group Analog (LGA) LGG~225 has been 
imaged with the Galaxy Evolution Explorer ({\it GALEX}) through its Far- and Near-UV 
wavebands. A significant fraction  of the group  members   appear to underwent  
recent/on-going interaction episodes that strongly disturbed overall galaxy morphology. UV-bright 
regions, sites of intense
star formation activity accompanied by intense dust extinction, mark the galaxy outskirts forming
irregular structures and tails.  Compared to the Local Group, LGG~225  
seems thus to  be experiencing a more intense and active evolutionary phase.
\end{abstract}

\section{Introduction}
\label{sec:1}
 
Poor groups of galaxies represent the defining aggregates of the so called ``field''.  Their 
importance comes from two facts: {\it i)} most of galaxies in the 
{\em Local} Universe is found in groups, rather than in the cluster 
environment \cite{Eke}. {\it ii)} The transition between galaxy properties 
typical of field and clusters happens just {\em at densities characteristic of poor groups}, 
thus suggesting the existence of re-processing mechanisms driven by the  
environment richness \cite{Lewis, Gomez}. Groups show a wide range of 
properties when adopting a multi-wavelength approach.  As far as X-ray 
emission is concerned, for instance, those dominated by ellipticals are 
rich in hot and diffuse intergalactic gas \cite{Mulch}, while in  spiral-dominated 
aggregates--- as is the case of our own Local Group--- a low X-ray emission 
comes from cold gas mainly confined  within fine structures (shells etc.) 
of single galactic sources (see \cite{Schim}).
Whether a link  exists between  these two different evolutionary outputs 
in a hierarchical cosmological framework is still matter of debate. 

In the group environment
we know that  galaxy encounters are extremely efficient in reshaping  member
morphology, being stellar velocity dispersion  within galaxies comparable to 
that of the group as a whole. At the same time, both the merger/accretion 
rate and the frequency of processes giving  rise to tidal dwarfs are still 
unknown. In this context, we are currently studying with {\it GALEX}
a sample of Local Groups Analogs which includes so far three systems, 
namely LGG~93, LGG~127 and LGG~225. 

We present here a preliminary
analysis of the  LGG~225 group composed of about 10 galaxies (e.g. \cite{Tul}). 
We imaged in the Near (NUV) and the Far (FUV) {\it GALEX} bands
the ultraviolet emission of NGC~3447, NGC~3447A,  NGC~3454, NGC~3455 
and UGC~6035 aiming at tracing the ongoing  star formation (SF) and its 
distribution across galaxies and their  intergalactic medium. In order to estimate 
the paramenters of these SF events we rely on the \cite{Buz1,Buz2} population synthesis 
models.

\section{The LGG~225 group: UV Morphology and Photometry}
\label{sec:2}

Table~1 provides a list  of spectroscopically confirmed group members,
 their optical classification \cite{Vauc} and 
 total {\it GALEX} FUV and NUV magnitudes. Our two {\it GALEX} pointings
 cover five members of LGG~225 sample, i.e. the bright part of its galaxy population.
The NUV and FUV images of  the five prevailing spirals are shown in the panels of Figure~1.  

\begin{table}
\centering
\caption{Main characteristics of the LGA~225}
\label{tab:1}       
 
\begin{tabular}{llcllcll}
\hline\noalign{\smallskip}
Name & Type & B & $V_r$ & NUV (AB) & FUV & (FUV - NUV) \\
\noalign{\smallskip}\hline\noalign{\smallskip}
NGC 3370 & SA(s)c & 12.28 & 1279 & 14.34 $\pm$ 0.01 \\
NGC 3443 & Sad & 13.70  & 1132 & 16.15 $\pm$ 0.01 & 
17.23  $\pm$ 0.02 & 1.08 $\pm$ 0.02\\
NGC 3447 & Mult. & & & 17.71 $\pm$ 0.02 \\
NGC 3447A & SAB(s)m pec & 13.10 & 1066 & 15.80 $\pm$ 0.01 & 
16.23 $\pm$ 0.01 & 0.43 $\pm$ 0.01\\
NGC 3447B & IB(s)m pec & & 1098 & & 16.43 $\pm$ 0.01 \\
NGC 3454 & SB(s)c? & 14.18 & 1101 & 18.01 $\pm$ 0.01 & 18.51 $\pm$ 0.03 & 
0.50 $\pm$ 0.03 \\
NGC 3455 &(R’)SAB(rs)b & 12.83 & 1102 & 14.719 $\pm$ 0.01 & \\
NGC 3507 & SB(s)b & 11.73 & 979 & 16.90 $\pm$ 0.03 & 18.21 $\pm$0.08 & 1.31 
$\pm$ 0.09 \\
UGC 5947 & Im pec. & 14.75 & 1251 & & & \\
UGC 6035 & IBm & 14.30 & 1072 & 16.25 $\pm$ 0.01 & 16.64 $\pm$ 0.01 & 0.39 
$\pm$ 0.01 \\
UGC 61121 & Sd? & 13.90 & 1033 & 16.89 $\pm$ 0.03 & 17.28 $\pm$ 0.05 & 0.39 
$\pm$ 0.06 \\
\noalign{\smallskip}\hline
\end{tabular}
\end{table}

Very little is reported in the literature about these galaxies. 
NGC~3447A and  its companion NGC~3447 are members 
of the KPG~255 pair in the \cite{Kara} catalogue with a radial velocity difference 
of about 30~km~s$^{-1}$. 

\begin{figure}

\includegraphics[height=3cm,width=0.49\hsize]{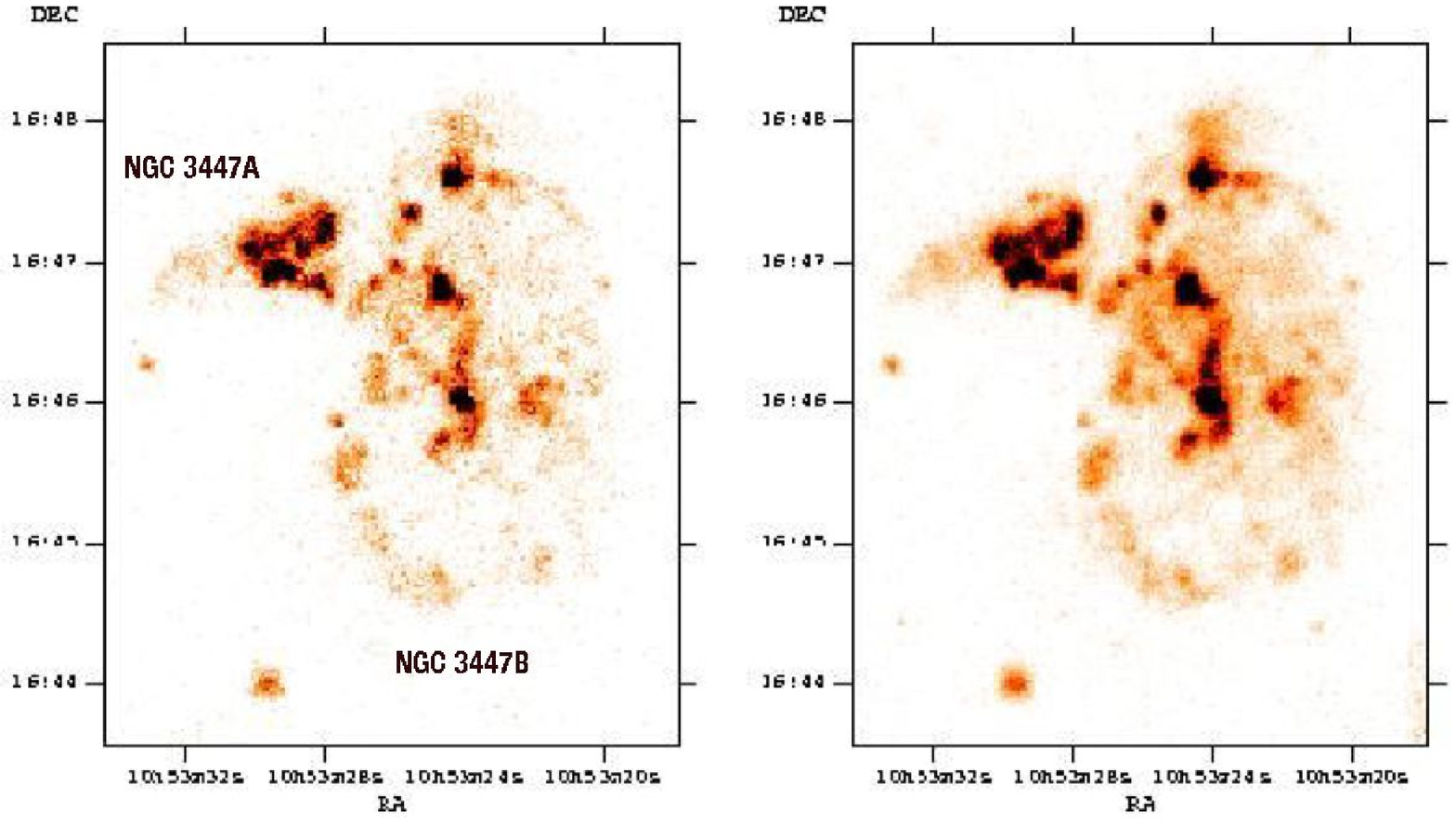}~ \includegraphics[height=3cm,width=0.49\hsize]{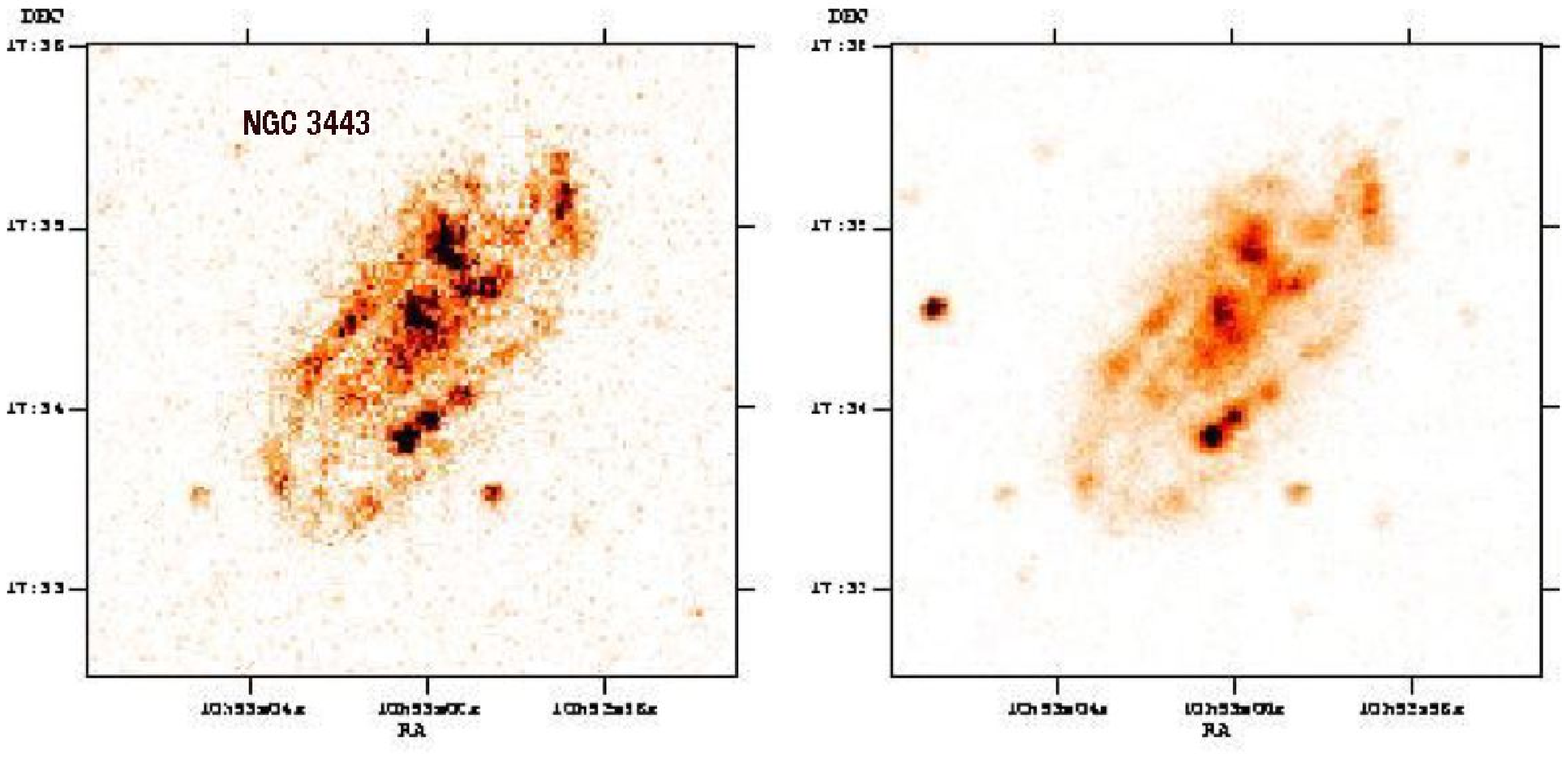}

\includegraphics[height=3cm,width=0.49\hsize]{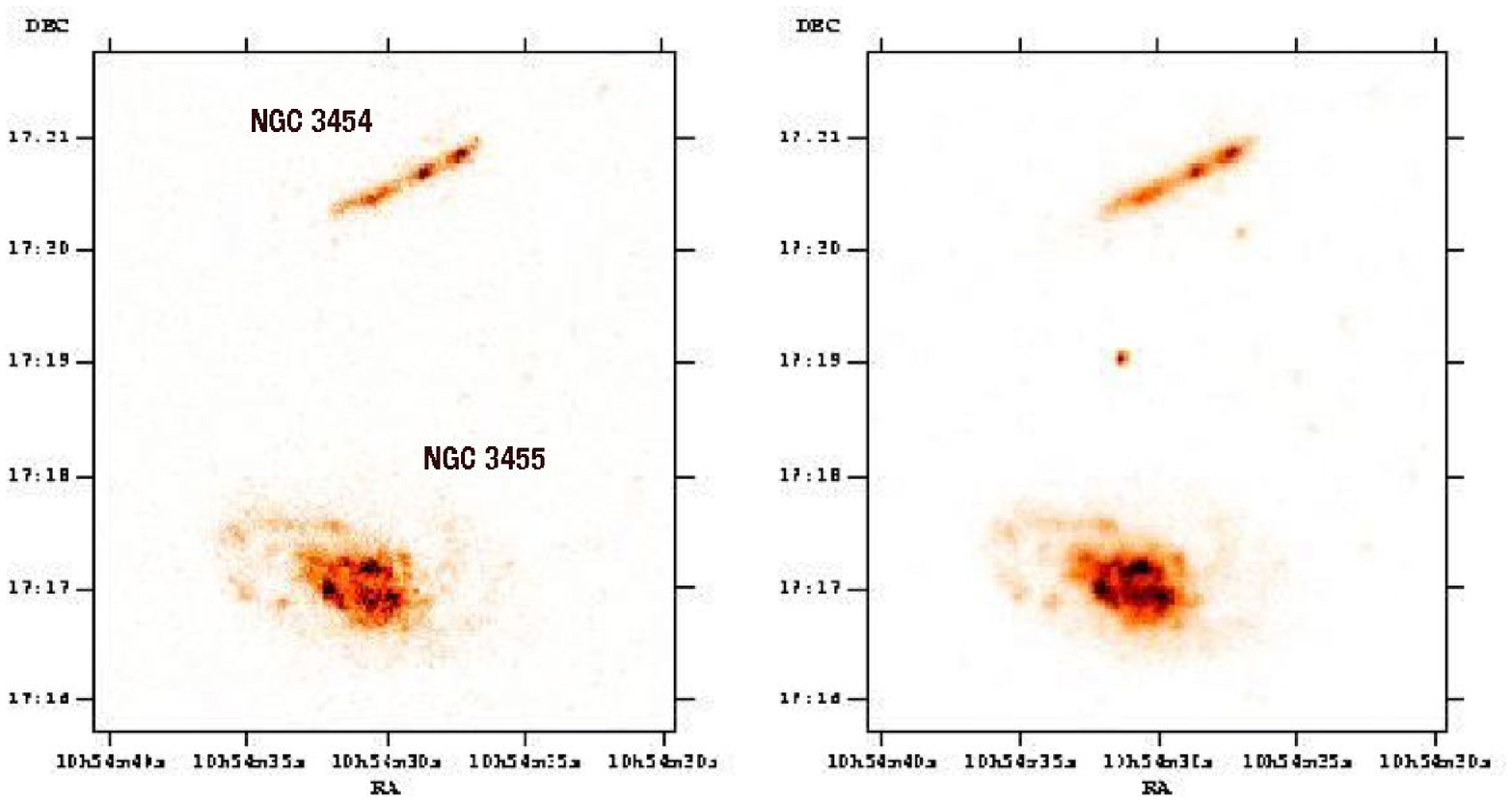}~ \includegraphics[height=3cm,width=0.49\hsize]{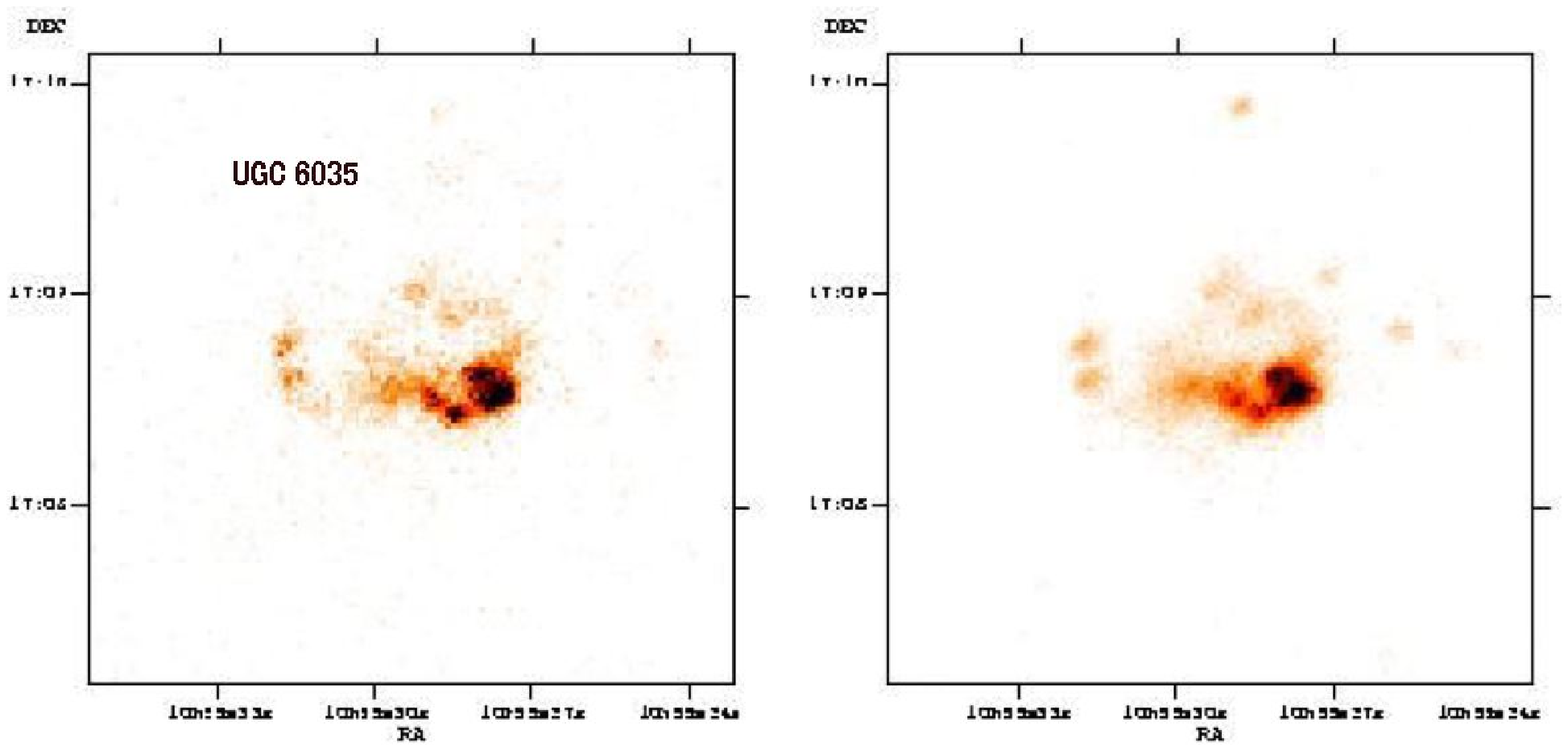}

\caption{{\it GALEX} FUV  and NUV   background-subtracted 
images of five members of the Local Group Analog LGG~225.} 
\label{fig:1}       
\end{figure}

The two galaxies appear strongly interacting. A bar is still visible in NGC 3447A, 
while both the underlying disk and the (multiple?) arms are tidally distorted. 
 Interaction is so strong that, for NGC~3447, it is even difficult 
to figure out the original morphological type (irregular or late-type spiral?).  As for 
NGC~3443, galaxy has no unambiguous  closeby companions, at least in projection. 
 Nevertheless, one can report that the northern half of the galaxy is completely 
different from the southern one:  clearly, a 2~D velocity field is necessary to 
understand the nature of this asymmetry. 

NGC~3454 and NGC~3455 are close both in projection and in the redshift space (Table~1). The first 
galaxy  displays a clean edge-on spiral morphology while the outer eastern arms of NGC~3455   
could be tidally disturbed.  Finally, both arms and disk of UGC 6035 are significantly 
asymmetric with star forming regions  confined in the western part of the galaxy 
nucleus. 
\begin{figure}
 \includegraphics[width=0.49\hsize, height=0.485\vsize]{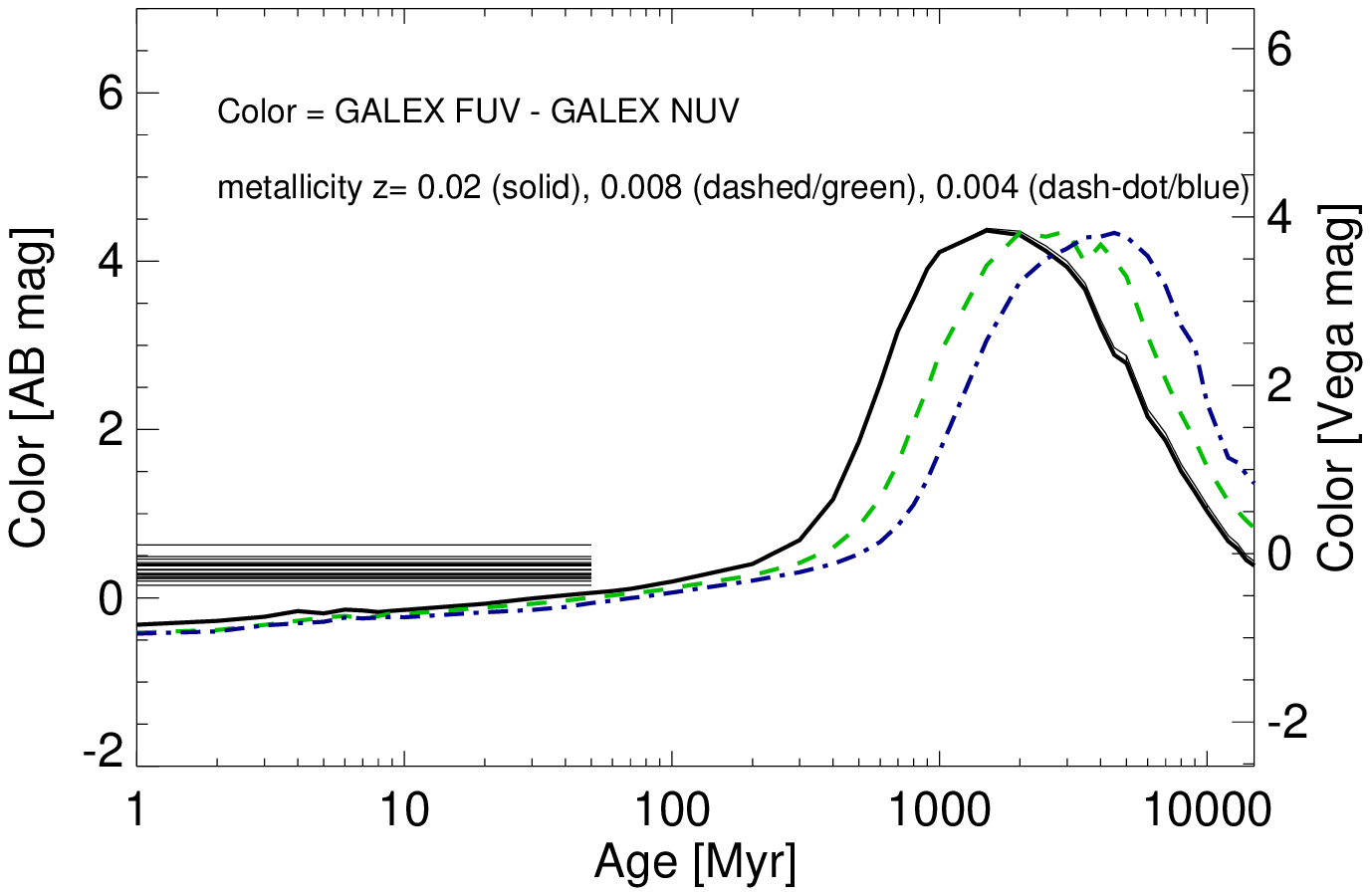}
 \hfill
 \includegraphics[width=0.47\hsize, height=0.49\vsize]{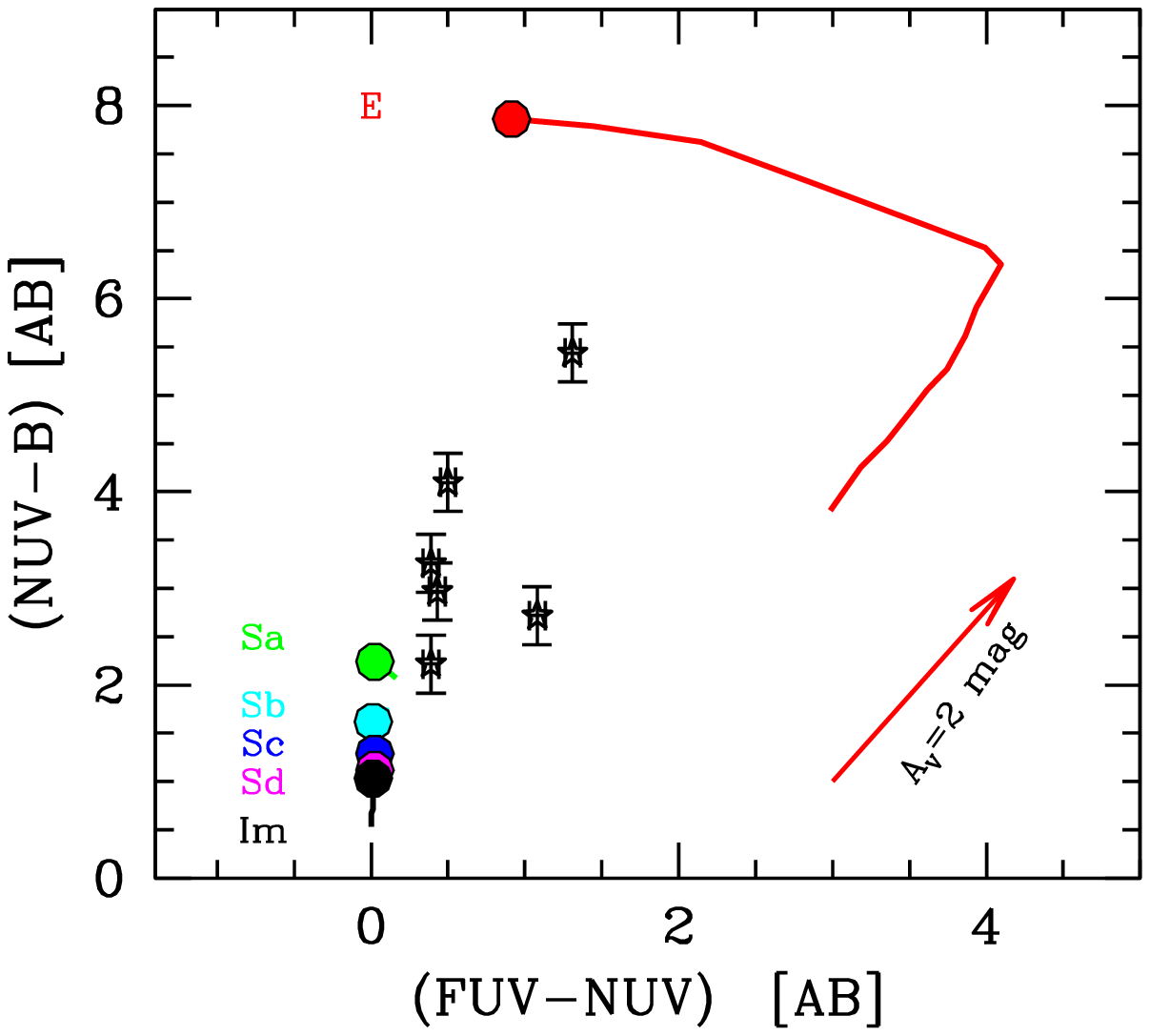}
\caption{{\it Left panel:} The GALEX FUV-NUV color for a Simple Stellar Population as a 
function of age (from the Padua  synthesis models) for three representative metallicity 
values. The measured colors 
of some SF regions of NGC 3447 are indicated by the tick line shelf on the left.
{\it Right panel:} Comparison of the integrated B vs. UV galaxy photometry of Table 1 
(star markers) with the Buzzoni (2002) galaxy templates of different morphology (from type 
E to Im, as labelled on the plot). Big dots for theoretical models refer to a 15 Gyr 
age, with back-in-time evolution traced down to 1 Gyr (see tails for the E and Im 
templates). The bottom right arrow is galaxy (internal) reddening for a 2~mag visual
extinction according to the Calzetti (1999) attenuation law. Note that observed 
galaxies seem to be  strongly affected by dust  in their UV colors, with a 
few 0.1 mag E(B-V) color excess.}
\end{figure}
\section{Preliminary UV analysis} 

When compared to the Local Group, LGG~225 seems to going through a much more 
active phase of evolution. Although the UV-optical colors of the innermost 
(often bar-like) bodies of the analyzed galaxies appear quite red  and are dominated 
by an old  stellar population, most of the systems show  outstanding UV-bright 
regions. 
Their measured (FUV-NUV) and (UV - optical) colors are consistent with the 
UV fluxes being dominated by moderately young (few hundred Myr or less) bursts of SF, 
accompanied by the presence of dust clouds (see Fig. 2).

Such UV bright, extended sources, especially those located in the tidal tails are
the ideal candidate to evolve towards  HI-rich, dwarf satellites (e.g. \cite{Neff,Hibba}), 
affecting the so-called ``missing satellite problem'' (cf. \cite{Krav}) and, more generally, 
actively contributing to the current evolution of the group. 



\printindex
\end{document}